\documentclass[
reprint,
superscriptaddress,
 amsmath,amssymb,
 aps,
prb,
]{revtex4-2}

\usepackage{chemformula}
\usepackage{siunitx}
\usepackage{graphicx}
\usepackage{dcolumn}
\usepackage{bm}
\usepackage{xcolor}
\usepackage[T1]{fontenc}
\usepackage{hyperref}

\definecolor{tugraz_red}{RGB}{226,0,26}

\newcommand{\elph}{\emph{el--ph}}
\newcommand{\Tc}{\textit{T}$_\text{c}^{\mkern2mu\text{onset}}$}
\begin{document}

\preprint{APS/123-QED}

\title{Inverse Isotope Effect in the Ternary Perovskite Hydride SrPdH/D$_{2.9}$: A Signature of Quantum Zero-Point Fluctuations} 

\author{Wencheng Lu}
\email{wlu@carnegiescience.edu}
\affiliation{Earth and Planets Laboratory, Carnegie Institution for Science, 5241 Broad Branch Road NW, Washington, DC 20015, USA}

\author{Mihir Sahoo}
\affiliation{Institute of Theoretical and Computational Physics, Graz University of Technology, NAWI Graz, 8010 Graz, Austria}

\author{Roman Lucrezi}
\affiliation{Department of Chemistry, Stockholm University, SE-10691 Stockholm, Sweden}

\author{Michael J. Hutcheon}
\affiliation{Enterprise Science Fund, Intellectual Ventures, 3150 139th Ave SE, Bellevue, WA, 98005, USA}

\author{Shubham Sinha}
\affiliation{Earth and Planets Laboratory, Carnegie Institution for Science, 5241 Broad Branch Road NW, Washington, DC 20015, USA}

\author{Pedro N. Ferreira}
\affiliation{Institute of Theoretical and Computational Physics, Graz University of Technology, NAWI Graz, 8010 Graz, Austria}

\author{Chris J. Pickard}
\affiliation{Department of Materials Science and Metallurgy, University of Cambridge, 27 Charles Babbage Road, Cambridge, CB3 0FS, UK}
\affiliation{Advanced Institute for Materials Research, Tohoku University, Sendai, 980-8577, Japan}

\author{Qiang Zhang}
\affiliation{Neutron Scattering Division, Oak Ridge National Laboratory, Oak Ridge, TN, USA}

\author{Matthew N. Julian}
\affiliation{Enterprise Science Fund, Intellectual Ventures, 3150 139th Ave SE, Bellevue, WA, 98005, USA}

\author{Rohit P. Prasankumar}
\affiliation{Enterprise Science Fund, Intellectual Ventures, 3150 139th Ave SE, Bellevue, WA, 98005, USA}

\author{Christoph Heil}
\email{christoph.heil@tugraz.at}
\affiliation{Institute of Theoretical and Computational Physics, Graz University of Technology, NAWI Graz, 8010 Graz, Austria}

\author{Timothy A. Strobel}
\email{tstrobel@carnegiescience.edu}
\affiliation{Earth and Planets Laboratory, Carnegie Institution for Science, 5241 Broad Branch Road NW, Washington, DC 20015, USA}

\date{\today}

\begin{abstract}
Guided by first-principles calculations, we demonstrate superconductivity in the ternary perovskite hydride SrPdH$_{3-x}$, synthesized at low pressure. Structural characterization via neutron diffraction reveals the near-stoichiometric composition SrPdD$_{2.9(2)}$ with 96\% deuterium site occupancy. Subsequent transport and magnetic susceptibility measurements establish onset superconducting transitions at $T_\text{c} = \SI{2.1}{K} $ (H) and $T_\text{c} = \SI{2.2}{K} $ (D), exhibiting an inverse isotope effect that our first-principles calculations attribute predominantly to quantum zero-point motion. The excellent agreement between theory and experiment with respect to thermodynamic stability and superconducting properties provides important validation for theory-guided superconductor discovery. This work establishes superconductivity in the perovskite hydride structural prototype---expanding the limited family of experimentally realized ternary hydride superconductors---and demonstrates the importance of quantum nuclear motion on the accurate theoretical treatment of low-pressure hydride superconductors.

\end{abstract}

\maketitle


The pursuit of high-$T_\text{c}$ superconductivity in hydride systems has emerged as an active frontier in condensed matter physics. To date, realizing superconductivity at elevated temperatures in hydrides has required the application of extreme pressures. Most known high-$T_\text{c}$ hydride superconductors fall within three well-established structural families: covalent hydrides (e.g., \ch{H3S}~\cite{drozdov2015conventional}), clathrate-type hydrides (e.g., \ch{LaH10}~\cite{drozdov2019superconductivity}, \ch{CaH6}~\cite{ma2022high}, and \ch{LaSc2H24}~\cite{song2025}), and molecular-type hydrides (e.g., \ch{BaH12}~\cite{chen2021synthesis}, \ch{BiH4}~\cite{shan2024molecular}). In an effort to lower the required synthetic pressure, substantial attention has been given to substitutional alloy hydrides~\cite{semenok2025}, such as \ch{(La,Y)H10}~\cite{semenok2021superconductivity}, \ch{(La,Ce)H9}~\cite{bi2022giant}, and \ch{(Y,Ce)H9}~\cite{chen2024synthesis}. However, the substitutional alloying strategy often entails trade-offs between structural stability and the superconducting transition temperature. These hydrogen-rich structures generally demand high external pressure to stabilize dense hydrogen frameworks, significantly limiting their feasibility for ambient-pressure applications. Therefore, achieving superconductivity at high temperatures under reduced or ambient pressure remains a central and unresolved challenge, making the search for alternative structural prototypes increasingly important. In recent studies, several ternary hydrides featuring unconventional structures have been successfully synthesized experimentally. These include the complex transition metal hydrides \ch{Li5MoH11}~\cite{meng2019superconductivity} and \ch{BaReH9}~\cite{muramatsu2015metallization}, the fluorite-type hydride \ch{LaBeH8}~\cite{song2023stoichiometric}, and the lanthanum-based borohydride \ch{LaB2H8}~\cite{song2024superconductivity}. However, these stoichiometric ternary compounds have also been found to be thermodynamically unstable at ambient pressure or to require high-pressure conditions to induce superconductivity. 

Other structural families may also offer potential for superconductivity. Metals naturally possess interstitial sites capable of trapping hydrogen, leading to the formation of lower-hydrogen-content metallic hydrides~\cite{Goldschmidt1967}. These systems often exhibit favorable electronic structures and can remain stable at relatively low pressures (e.g., PdH with $T_\text{c}$ $\approx$ \SI{9}{K}~\cite{stritzker1972superconductivity}). In such hydrides, hydrogen atoms are embedded within an electron-rich metallic matrix, enabling superconductivity without requiring extreme hydrogen ionization or ultrahigh pressure. Recent theoretical studies have also identified other candidates with high-$T_\text{c}$s under ambient or near-ambient pressures. These include complex transition-metal hydrides such as \ch{Mg2IrH6} (65--170~\si{K})~\cite{dolui2024feasible,sanna2024prediction,zheng2024prediction,Hansen_2024}, perovskite hydrides exemplified by \ch{KInH3} (\SI{73}{K})~\cite{cerqueira2024searching}, and anti-perovskite hydrides like \ch{ZnHAl3} (\SI{80}{K})~\cite{liu2024high}. Recent low-pressure studies in the Mg--Pt--H system revealed superconductivity in the novel complex hydride \ch{Mg4Pt3H6}~\cite{Lu_2025}.

\begin{figure}[t!]
\begin{center}
\includegraphics[width=0.47\textwidth]{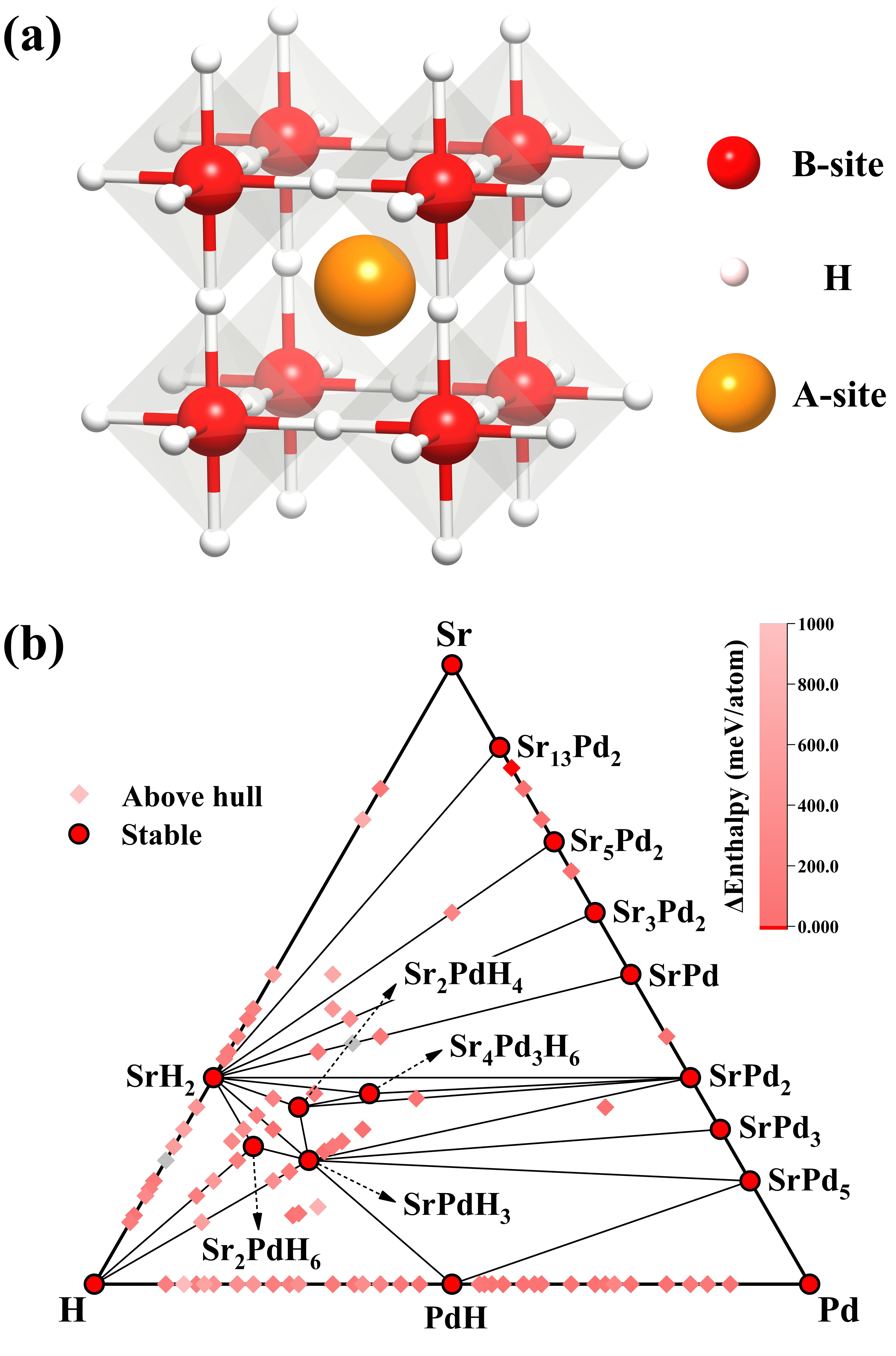}\\[5pt] 
\caption{\textit{\textbf{a}} Crystal structure of perovskite hydrides, where the A-site, B-site and H atoms are shown as orange, red, and white spheres. In \ch{SrPdH3}, Pd and Sr occupy the B-site and A-site, respectively. \textit{\textbf{b}} Calculated ternary phase diagram of the Sr--Pd--H system at ambient pressure. Structures are represented by rhombuses and colored according to their distance from the convex hull; thermodynamically stable structures are indicated by red solid circles.}
\label{fig:Fig1}
\vspace{-0.4 cm} 
\end{center}
\end{figure}

Perovskite hydrides, characterized by a three-dimensional framework with A- and B-site cations and corner-sharing hydride octahedra (Fig.~\ref{fig:Fig1}(a)), are fundamentally different to other high-$T_\text{c}$ hydride prototypes. Unlike covalent or molecular hydrides, these materials exhibit hydridic bonding, which enhances metallicity and may facilitate superconducting pairing. Given their structural versatility, promising low-pressure thermodynamic stability, and high predicted $T_\text{c}$s, perovskite-type hydrides stand out as a compelling subclass in the search for practical high-$T_\text{c}$ superconductors. Numerous compounds have been proposed as promising candidates exhibiting high $T_\text{c}$s at ambient conditions. Examples include \ch{RbPH3}~\cite{DANGIC2025100043}, \ch{KMH3} with M = Al (\SI{52}{K}), Cd (\SI{23.4}{K})~\cite{wines2024data}, \cite{cerqueira2024sampling}),
\ch{MScH3}~\cite{quan2025prediction} with M = K (\SI{40.4}{K}), Rb (\SI{32.9}{K}), Cs (\SI{18.6}{K}),
\ch{SrMH3}~\cite{li2024theoretical} with M = Au (\SI{132}{K}), Tc (\SI{69}{K}), Zn (\SI{107}{K}), \ch{Al4H} (\SI{54}{K})~\cite{he2023phonon}, \ch{PbMH3}~\cite{cerqueira2024searching} with M = Hg (\SI{25.8}{K}), Os (\SI{23.2}{K}), \ch{AlHgH3} (\SI{28.4}{K})~\cite{cerqueira2024searching}. Furthermore, Du \textit{et al.} proposed five perovskite hydrides with $T_\text{c}$s exceeding \SI{120}{K} under 10 GPa~\cite{du2024high}, \ch{KGaH3} (\SI{146}{K}), \ch{RbInH3} (\SI{130}{K}), \ch{CsInH3} (\SI{153}{K}), \ch{RbTlH3} (\SI{170}{K}), and \ch{CsTlH3} (\SI{163}{K}). Despite the growing body of computational studies exploring superconductivity in this class of hydrides, the potential of this structure type remains unverified by experiment. The absence of direct experimental confirmation leaves a critical gap between theoretical predictions and practical realization, underscoring the need for targeted synthesis and characterization efforts.

In this work, we report the synthesis and characterization of SrPdH$_{3-x}$, a cubic perovskite superconductor that provides new insights into the physics of hydride superconductors. Using high-throughput computational screening, we identified this material as thermodynamically stable under ambient pressure. Subsequently, we were able to synthesize SrPdH$_{3-x}$ under moderate conditions (\SI{560}{bar}, \SI{475}{^\circ C}), achieving the near-stoichiometric composition SrPdD$_{2.9(2)}$, as confirmed by neutron diffraction. Notably, our electrical transport and magnetic susceptibility measurements reveal an inverse isotope effect: deuterated samples exhibit slightly higher superconducting transition temperatures compared to their hydrogenated counterparts. Through state-of-the-art first-principles calculations combining density functional perturbation theory (DFPT)~\cite{baroni2001phonons} with the stochastic self-consistent harmonic approximation (SSCHA)~\cite{Werthamer_1970,Monacelli_2021}, we demonstrate that this behavior arises from quantum nuclear effects. Our results provide the first quantitatively accurate theoretical reproduction of the inverse isotope effect in a ternary hydride superconductor, revealing fundamental physics that will be important for understanding and designing ambient-pressure superconducting hydrides.

As illustrated in the thermodynamic convex hull (Fig.~\ref{fig:Fig1}(b)), our high-throughput calculations identify four thermodynamically stable ternary hydrides in the Pd--Sr--H system at ambient pressure: \ch{SrPdH3}, \ch{Sr2PdH4}, \ch{Sr2PdH6}, and \ch{Sr4Pd3H6}. 
Among these, our initial coarse screening identified \ch{SrPdH3} to be dynamically stable and to exhibit electron-phonon coupling, later refined to give a $T_\text{c}$ of approximately \SI{2.8}{K} through high-accuracy calculations described below.

To the best of our knowledge, \ch{Sr2PdH6} and \ch{Sr4Pd3H6} remain unreported in prior experimental or theoretical studies. Their absence in experiments may be attributed to finite temperature effects or large kinetic barriers, potentially requiring alternative synthetic approaches.
Synthesis of \ch{Sr2PdH4}, which adopts an orthorhombic crystal structure and exhibits insulating behavior, was reported in previous studies~\cite{stanitski1972ternary,olofsson1999novel}, including both hydrogenated and deuterated forms.

SrPdH$_{3-x}$ was previously synthesized by Bronger and Ridder in the 1990s~\cite{bronger1994synthese}, though our theoretical prediction and synthesis were conducted independently of this earlier work. This phase crystallizes in the perovskite-type framework, where Sr atoms occupy the A-sites, and Pd atoms located at the B-sites are octahedrally coordinated by six hydrogen atoms that reside at the 3$d$ Wyckoff positions and exhibit 4/\textit{mm.m} site symmetry. Previous neutron diffraction measurements on samples synthesized at \SI{5}{bar} indicated a composition of SrPdH$_{2.7}$, and the material was described as having a metallic copper-like appearance, though no physical properties were reported~\cite{bronger1994synthese}. Later, Yagyu \textit{et al.} prepared SrPdH$_{3-x}$ with unidentified impurities using \ch{CaH2} as the hydrogen source at $\sim$\SI{5}{bar} and room temperature~\cite{yagyu2013synthesis}. Based on magnetic susceptibility measurements, no superconducting transition was observed above \SI{2}{K}. 

\begin{figure}[t!]
\begin{center}
\includegraphics[width=0.48\textwidth]{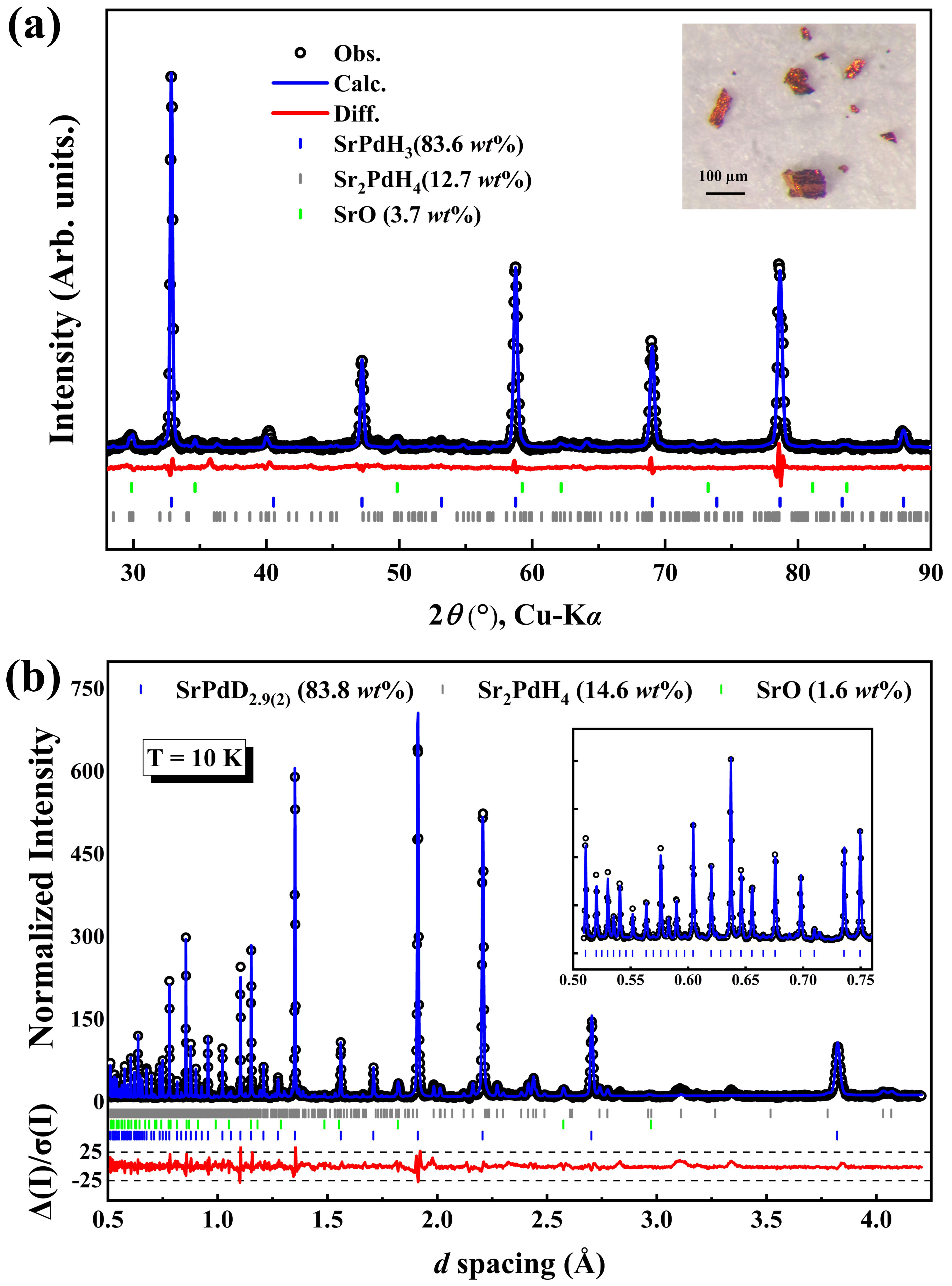}\\[5pt] 
\caption{\textit{\textbf{a}} XRD pattern (black points) for hydrogenated samples (measured at room temperature) with Rietveld refinement (\textit{R}$_{wp}$ = 4.89\%) of the predicted \textit{Pm}$\Bar{3}$\textit{m} \ch{SrPdH3} structure (blue line). The red line shows the Rietveld residual and vertical ticks marks indicate allowed Bragg reflections. Impurity peaks from \ch{SrO} and \ch{Sr2PdH4} are represented by green and grey ticks, respectively. Inset: photographs of the sample after hydrogenation. \textit{\textbf{b}}  TOF-neutron diffraction pattern for deuterated sample at \SI{10}{K} (black points) and Rietveld refinement (blue line) with \textit{R}$_{wp}$ of 8.17\%. The red line shows the Rietveld residual and the vertical ticks mark the positions of nuclear peaks for SrPdD$_{2.9(2)}$ (blue), \ch{SrO} (green) and \ch{Sr2PdD4} (grey).}
\label{fig:Fig2}
\vspace{-0.4cm} 
\end{center}
\end{figure}

Following our independent predictions, we attempted synthesis of \ch{SrPdH3} by hydrogenating SrPd~\cite{iandelli1974europium} (see Fig. S1) under relatively moderate conditions of \SI{560}{bar} and 475~$^\circ$C (More details on the experimental and theoretical methods in Supplemental Information and Refs.~\cite{QE1, giannozzi2017advanced, QE3,Garrity2014,Evans2024,Merchant2023,DFT,Kohn-Sham,DFPT,PBE,ONCV1, ONCV2,MPsampling, toby2013gsas, MP-smearing, Shapeev2015-MTP,Novikov_2021,lucrezi2023quantum,lucrezi2024,lee2023electron,ferreira2024}). After hydrogenation/ deuteration, the copper-colored product remains stable at ambient pressure (Fig.\ref{fig:Fig2}(a), inset), but decomposes in air and must be handled in an inert environment. 
In order to verify the crystal structure of the synthesized compound, powder XRD measurements were conducted at ambient pressure, as shown in Fig.~\ref{fig:Fig2}(a). Rietveld analysis of the diffraction patterns indicate the formation of the SrPdH$_{3-x}$ perovskite framework (\textit{Pm}$\Bar{3}$\textit{m}) with minor \ch{Sr2PdH4} and SrO impurities. The experimental lattice parameter $a$ = 3.839(3) \AA~ is in excellent agreement with the predicted theoretical value $a$ = 3.848 \AA~---~a difference of only 0.23\%.

To determine the hydrogen content of the synthesized material, deuterated samples were prepared under the same conditions and analyzed using neutron powder diffraction. Rietveld refinement of the diffraction pattern obtained at 10 K (Fig.~\ref{fig:Fig2}(b), with $a$ = 3.821(8) \AA) yields the composition SrPdD$_{2.9(2)}$, showing higher deuterium content than reported previously~\cite{bronger1994synthese}. This deviation is presumably due to the higher synthesis pressure used in this study (\SI{560}{bar}) compared to the \SI{5}{bar} pressure used in earlier work.

\begin{figure}[t!]
\begin{center}
\includegraphics[width=0.48\textwidth]{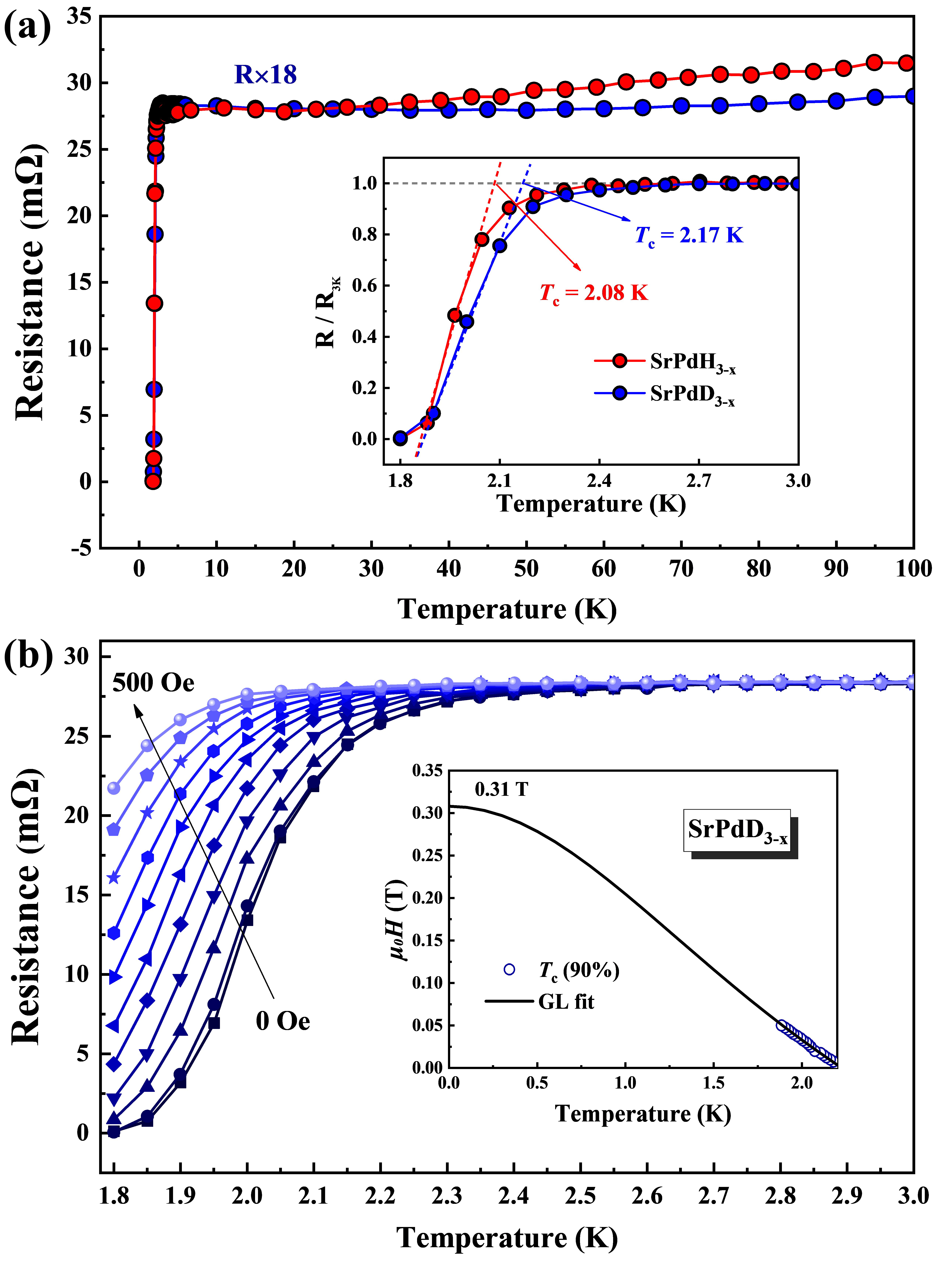}\\[5pt] 
\caption{\textit{\textbf{a}} Electrical transport measurements of the synthesized hydride and deuteride under ambient pressure. Inset show the \Tc{} by extrapolating the linear portions of the resistance curves. \textit{\textbf{b}} Temperature dependence of the electrical resistance for deuteride under applied magnetic fields up to \SI{500}{Oe}, measured in steps of \SI{50}{Oe}. The inset shows the GL fit and estimation of the upper critical field.}
\label{fig:Fig3}
\vspace{-0.4cm} 
\end{center}
\end{figure}

The potential for superconductivity was investigated by performing electrical transport measurements with the assistance of a Quantum Design Physical Property Measurement System (PPMS, Model 6000). Representative temperature-dependent electrical transport data are presented in Fig.~\ref{fig:Fig3}(a). Samples exhibit metallic conductivity with a clear superconducting transition characterized by a sharp drop to zero resistance for both the hydrogenated and deuterated samples.
The superconducting onset temperature, \Tc{}, established by extrapolating the linear portions of the resistance curves immediately above and below the transition, was determined to be \SI{2.1}{K} for the hydrogenated sample. Strikingly, deuterated samples exhibit a slightly enhanced \Tc{} of \SI{2.2}{K}. This anomalous increase, opposite to the suppression anticipated within the conventional BCS framework~\cite{bardeen1957theory}, constitutes clear evidence of an \textit{inverse isotope effect}. This counterintuitive behavior highlights anomalous aspects of the superconducting mechanism in these hydrides and will be revisited in detail below.

To further probe superconductivity, magnetic fields up to \SI{0.05}{T} were applied during electrical transport measurements to investigate the upper critical field (as shown for the deuterated sample in Fig.~\ref{fig:Fig3}(b) and hydrogenated sample in Fig. S3). The observed resistance drop exhibits systematic suppression with increasing field, further confirming the superconducting nature of the transition. To estimate the upper critical field, $H_{\text{c}2}$, the observed field-dependent transition temperatures were fitted to a modified Ginzburg–Landau (GL) expression, $H_{\text{c}2}(T) = H_{\text{c}2}(0)(1-(T/T_\text{c})^2)$, where $H_{\text{c}2}(T)$ is the upper critical field at temperature $T$, using the criterion of 90\% of the normal-state resistance at \SI{3}{K} to define $T_\text{c}$. From this analysis, we estimate a zero-temperature value of $\mu_0H_{\text{c}2}(0) = \SI{0.31}{T}$ of deuterated sample.

To further confirm the superconducting nature of the transition observed by electrical transport, temperature-dependent AC magnetic susceptibility measurements were performed on H- and D-bearing samples using the ACSM-I option of the PPMS (see Supplementary Information). In both cases, samples clearly show magnetic field expulsion with a sharp drop in the real component of the susceptibility ($\chi^\prime$), signaling a diamagnetic transition, as well as energy dissipation with a rise in the imaginary component of the susceptibility ($\chi^{\prime\prime}$). These observations confirm the superconducting nature of the samples and corroborate the observation from transport that \Tc{} for SrPdD$_{3-x}$ is higher than that of SrPdH$_{3-x}$ (Fig. S4). 

To obtain a comprehensive theoretical understanding of the superconducting properties of these compounds, we performed additional high-accuracy calculations. We computed electron--phonon (\elph{}) matrix elements on dense grids using DFPT as implemented in Quantum ESPRESSO~\cite{giannozzi2017advanced}, followed by Wannier interpolation via EPW to achieve even denser Brillouin zone (BZ) sampling~\cite{lee2023electron,lucrezi2024full}. This approach ensures full convergence of the BZ integrals required for computing the Eliashberg spectral function $\alpha^2F(\omega)$~\cite{migdal1958,eliashberg1960}, which serves as the primary input for the Eliashberg equations to estimate $T_\text{c}$ from first principles~\cite{kogler2025isome,Pellegrini2024}. The spectral function also enables calculation of two key parameters that characterize conventional superconductors~\cite{AllenDynes1975}: the total \elph{} coupling strength $\lambda = \int  \frac{d\omega }{\omega} \alpha^2 F(\omega)$ and the \elph{}-weighted averaged frequency $\omega_\text{log} = \text{exp}\left[ \frac{2}{\lambda} \int d\omega \frac{\text{ln}(\omega)}{\omega} \alpha^2 F(\omega) \right]$.

Our calculations yield $\lambda = 0.4$ for both hydrogen and deuterium compounds, but reveal a larger \elph{}-weighted averaged frequency $\omega_\text{log}$ of \SI{57}{meV} for \ch{SrPdH3} compared to \SI{43}{meV} for \ch{SrPdD3}, reflecting the higher vibrational frequencies of the lighter hydrogen atoms. Consequently, \ch{SrPdH3} exhibits a slightly higher $T_\text{c}$ of \SI{2.8}{K} compared to \SI{2.1}{K} for \ch{SrPdD3}, using the McMillan--Allen--Dynes formula~\cite{AllenDynes1975} and the commonly adopted value $\mu^*_\text{AD} = 0.1$ for the Morel-–Anderson Coulomb pseudopotential~\cite{morel1962a} in place of a full first-principles evaluation~\cite{kogler2025isome}. While these numerical $T_\text{c}$ values closely match experimental results, they fail to reproduce the experimentally observed inverse isotope effect.

This is to be expected, as the above described workflow neglects quantum ionic anharmonic effects, which can be significant for compounds containing very light elements. In order to account for these effects, we employed the stochastic self-consistent harmonic approximation (SSCHA) with \textit{ab initio}-trained machine learning potentials~\cite{Shapeev2015-MTP,Novikov_2021}, a cutting-edge approach that has only recently become computationally feasible for complex superconducting systems. In particular, we developed highly accurate interatomic potentials from our DFT data and utilized them to fully converge the SSCHA calculations, following the methodology outlined in Refs.~\cite{lucrezi2023quantum,lucrezi2024}. 

These calculations reveal that \ch{SrPdH3} requires a larger equilibrium volume than \ch{SrPdD3}, which can be attributed to the larger zero-point motion displacements of hydrogen compared to deuterium. Expressing the same point from a different perspective: when SSCHA calculations are performed on structures relaxed within the Born--Oppenheimer approximation, they generate internal pressures of \SI{2.1}{GPa} for \ch{SrPdH3} and \SI{1.6}{GPa} for \ch{SrPdD3}.

\begin{figure}[t!]
\begin{center}
\includegraphics[width=0.48\textwidth]{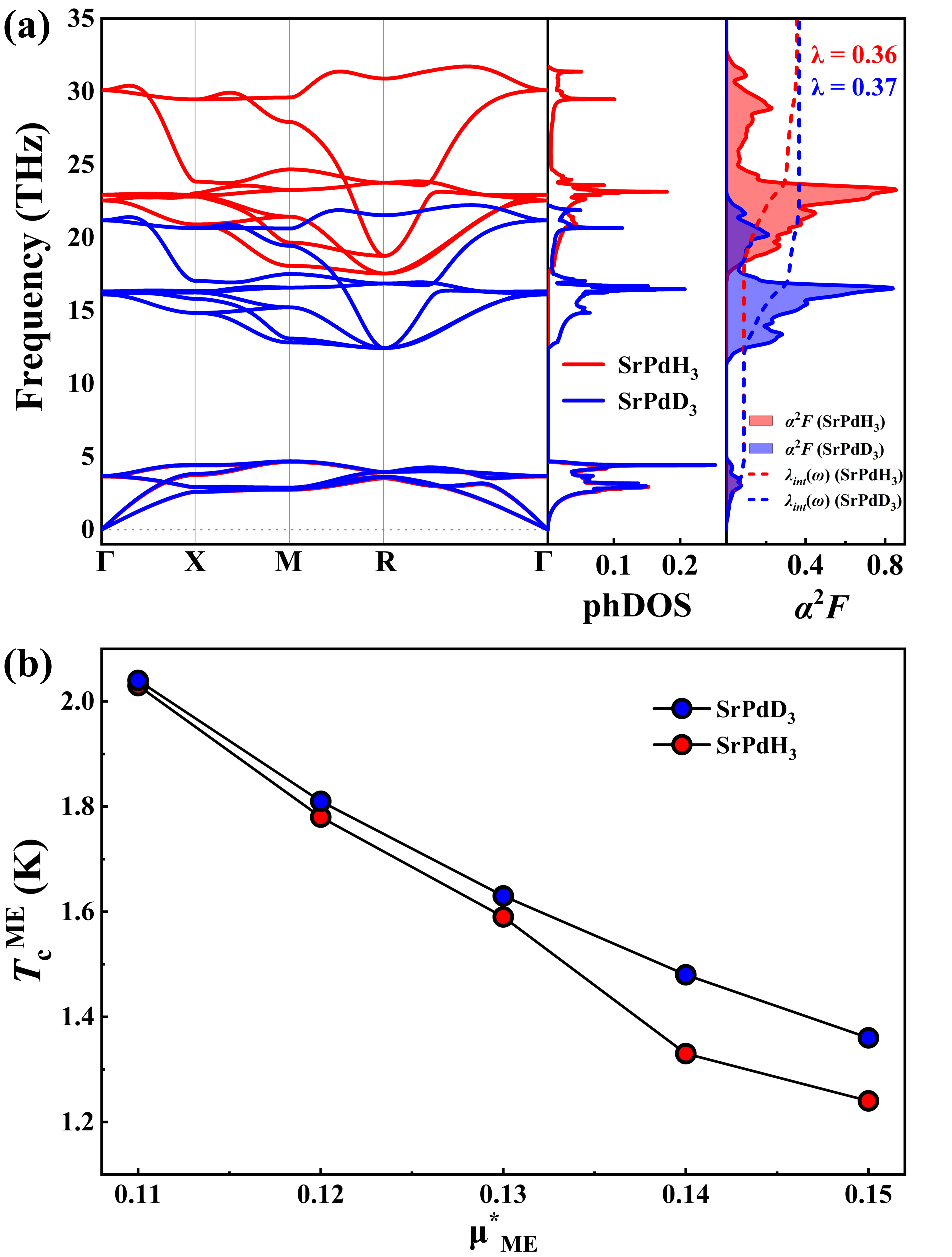}\\[5pt] 
\caption{\textit{\textbf{a}} Anharmonic phonon dispersion, phonon DOS, Eliashberg spectral function $\alpha^2F(\omega)$, and cumulative \elph{} coupling $\lambda(\omega)$. \textbf{b} Critical temperatures determined via Migdal-Eliashberg theory as a function of screened Coulomb pseudopotential $\mu^*$.}
\label{fig:Fig4}
\vspace{-0.4cm} 
\end{center}
\end{figure}

The volume expansion induces sizable shifts in the phonon spectrum, as can be appreciated in Fig.~\ref{fig:Fig4}(a), which reduce the superconducting parameters of \ch{SrPdH3} to $\lambda = 0.36$ and $\omega_\text{log} = \SI{55}{meV}$, compared with $\lambda = 0.37$ and $\omega_\text{log} = \SI{41}{meV}$ for \ch{SrPdD3}. Eliashberg calculations using IsoME~\cite{kogler2025isome} give $T_\text{c}$ values between 1.2 and \SI{2.0}{K}, depending on $\mu^*$ [Fig.~\ref{fig:Fig4}(b)], in excellent agreement with experiment. Moreover, for $\mu^* > 0.1$, the calculated $T_\text{c}$ of \ch{SrPdD3} consistently exceeds that of \ch{SrPdH3}, thereby reproducing the experimentally observed inverse isotope effect. The measured isotope coefficient of $\alpha \approx -0.07$ is accurately reproduced within the numerical precision of our calculations, confirming that the anomalous isotope dependence originates from the anharmonic volume expansion.

Further analysis reveals that anharmonicity in its strictest sense (non-parabolicity of the potential energy surface) and phonon--phonon interactions have minimal impact on the phonon and \elph{} properties. This highlights a fundamentally different mechanism for the inverse isotope effect compared to the extensively studied binary \ch{PdH(D)} system~\cite{stritzker1972superconductivity,schirber1974concentration}. While Errea \textit{et al.}~\cite{errea2013first} demonstrated that anharmonic effects in PdH arise from double-well potentials that stabilize otherwise unstable harmonic phonons, the situation in \ch{SrPdH3} is markedly different. Here, the harmonic phonons are already stable and yield excellent agreement with experimental $T_\text{c}$ values. Instead, the inverse isotope effect emerges predominantly from zero-point motion differences between H and D isotopes, which drive volume adjustments that shift the phonon dispersion without requiring phonon--phonon interactions or potential anharmonicity.

Importantly, we propose that this quantum zero-point mechanism becomes progressively weaker at higher pressures, where increased structural rigidity suppresses volume expansion effects~\cite{lucrezi2023quantum}. This pressure dependence provides a natural explanation for why high-pressure superconductors such as \ch{H3S} and \ch{LaH10} exhibit conventional isotope effects rather than the inverse behavior observed here~\cite{drozdov2015conventional,drozdov2019superconductivity}, although structure-specific anharmonic effects are also relevant. These insights are therefore particularly significant for the ongoing search for higher-$T_\text{c}$ ambient and low-pressure hydride superconductors, where accurate theoretical predictions require explicit treatment of quantum nuclear motion to capture isotope-dependent volume effects.

Finally, we discuss potential impacts related to compositional variability. While H- and D-bearing samples were synthesized under the same experimental conditions, we cannot explicitly rule out minor compositional variations that might contribute to differences in observed $T_\text{c}$s. Our calculations within the Generalized Quasichemical Approximation (GQCA)~\cite{ferreira2024} incorporating hydrogen and deuterium vacancies (see SI for details) indicate that the density of states around the Fermi level remains almost unchanged for low vacancy concentrations (Fig. S5), and $T_\text{c}$ is therefore expected to remain approximately constant across small changes in composition. Across all of our experiments, the observed $T_\text{c}$ onset temperature was always higher for D samples than for H samples. These observations suggest that the systematic shift between H and D is robust with an origin rooted in zero-point nuclear quantum motion. Indeed, recent computational studies highlight the importance of anharmonic effects within the perovskite hydride structural prototype \cite{DANGIC2025100043}. SrPdD/H$_{3-x}$ thus appears to be the second known example of the inverse isotope effect in hydride superconductors. While previous measurements on SrPdH$_{3-x}$ did not observe a superconducting transition \cite{yagyu2013synthesis}, susceptibility measurements were only performed to \SI{2}{K}, which may not be sufficiently low temperature to observe an unambiguous transition near the onset temperature for the hydride while screening.

In summary, we establish superconductivity in the perovskite hydride structural prototype using SrPdH/D$_{3-x}$, which represents a promising framework for hydride superconductors. Experimental measurements and theoretical calculations show that deuterated samples exhibit an anomalously higher \Tc{} (2.2 K) compared with hydrogenated samples (2.1 K), showing clear evidence of inverse-isotope-dependent behavior. By highlighting the essential role of quantum nuclear effects, our results broaden the scope for exploring anomalous superconductivity in complex hydrides and provide a framework for extending research possibilities to other low-pressure, hydrogen-rich materials.

\begin{acknowledgments}
This work used the Spallation Neutron Source (SNS), a DOE Office of Science User Facility operated by the Oak Ridge National Laboratory. The beam time was allocated to Powgen BL-11A under proposal number IPTS-32521.1. The research was supported by the Enterprise Science Fund of Intellectual Ventures.
Computations were performed on the lCluster at TU Graz and the Austrian Scientific Computing clusters VSC4 and VSC5. R.L. acknowledges the Carl Tryggers Stiftelse för Vetenskaplig Forskning (CTS 23: 2934). PNF acknowledges support from the Austrian Science Fund (FWF) under project DOI 10.55776/ESP8588124.
\end{acknowledgments}


\bibliography{ref}

@article{du2024high,
  title={{High-Temperature Superconductivity in Perovskite Hydride Below \SI{10}{GPa}}},
  author={Du, Mingyang and Huang, Hongyu and Zhang, Zihan and Wang, Min and Song, Hao and Duan, Defang and Cui, Tian},
  journal={Adv. Sci.},
  volume={11},
  number={42},
  pages={2408370},
  year={2024},
  doi = {https://doi.org/10.1002/advs.202408370},
  publisher={Wiley Online Library}
}

@article{drozdov2015conventional,
  title={{Conventional superconductivity at 203 kelvin at high pressures in the sulfur hydride system}},
  author={Drozdov, AP and Eremets, MI and Troyan, IA and Ksenofontov, Vadim and Shylin, Sergii I},
  journal={Nature},
  volume={525},
  number={7567},
  pages={73--76},
  year={2015},
  doi = {https://doi.org/10.1038/nature14964},
  publisher={Nature Publishing Group UK London}
}

@misc{song2025,
  title={{Room-Temperature Superconductivity at \SI{298}{K} in Ternary {La-Sc-H} System at High-pressure Conditions}}, 
  author={Yinggang Song and Chuanheng Ma and Hongbo Wang and Mi Zhou and Yanpeng Qi and Weizheng Cao and Shourui Li and Hanyu Liu and Guangtao Liu and Yanming Ma},
  year={2025},
  eprint={2510.01273},
  archivePrefix={arXiv},
  primaryClass={cond-mat.supr-con},
  url={https://arxiv.org/abs/2510.01273}, 
}

@article{DANGIC2025100043,
  title = {{Ambient pressure high temperature superconductivity in RbPH3 facilitated by ionic anharmonicity}},
  journal = {Comput. Mater. Today},
  volume = {8},
  pages = {100043},
  year = {2025},
  issn = {2950-4635},
  doi = {https://doi.org/10.1016/j.commt.2025.100043},
  author = {Đorđe Dangić and Yue-Wen Fang and Tiago F.T. Cerqueira and Antonio Sanna and Trinidad Novoa and Hao Gao and Miguel A.L. Marques and Ion Errea},
}

@article{iandelli1974europium,
  title={{The europium-palladium system}},
  author={Iandelli, A and Palenzona, A},
  journal={J. Less-Common Met.s},
  volume={38},
  number={1},
  pages={1--7},
  year={1974},
  publisher={Elsevier},
  doi={https://doi.org/10.1016/0022-5088(74)90197-0},
}

@article{drozdov2019superconductivity,
  title={{Superconductivity at \SI{250}{K} in lanthanum hydride under high pressures}},
  author={Drozdov, AP and Kong, PP and Minkov, VS and Besedin, SP and Kuzovnikov, MA and Mozaffari, S and Balicas, L and Balakirev, Fedor Fedorovich and Graf, DE and Prakapenka, VB and others},
  journal={Nature},
  volume={569},
  number={7757},
  pages={528--531},
  year={2019},
  doi = {https://doi.org/10.1038/s41586-019-1201-8},
  publisher={Nature Publishing Group UK London}
}

@article{ma2022high,
  title={{High-temperature superconducting phase in clathrate calcium hydride \ch{CaH6} up to \SI{215}{K} at a pressure of \SI{172}{GPa}}},
  author={Ma, Liang and Wang, Kui and Xie, Yu and Yang, Xin and Wang, Yingying and Zhou, Mi and Liu, Hanyu and Yu, Xiaohui and Zhao, Yongsheng and Wang, Hongbo and others},
  journal={Phys. Rev. Lett.},
  volume={128},
  number={16},
  pages={167001},
  year={2022},
  doi = { https://doi.org/10.1103/PhysRevLett.128.167001},
  publisher={APS}
}

@article{chen2021synthesis,
  title={{Synthesis of molecular metallic barium superhydride: pseudocubic \ch{BaH12}}},
  author={Chen, Wuhao and Semenok, Dmitrii V and Kvashnin, Alexander G and Huang, Xiaoli and Kruglov, Ivan A and Galasso, Michele and Song, Hao and Duan, Defang and Goncharov, Alexander F and Prakapenka, Vitali B and others},
  journal={Nature commun.},
  volume={12},
  number={1},
  pages={273},
  year={2021},
  doi = {https://doi.org/10.1038/s41467-020-20103-5},
  publisher={Nature Publishing Group UK London}
}

@article{shan2024molecular,
  title={{Molecular Hydride Superconductor \ch{BiH4} with $T_c$ up to \SI{91}{K} at \SI{170}{GPa}}},
  author={Shan, Pengfei and Ma, Liang and Yang, Xin and Li, Mei and Liu, Ziyi and Hou, Jun and Jiang, Sheng and Zhang, LiLi and Shi, Lifen and Yang, Pengtao and others},
  journal={J. Am. Chem. Soc.},
  volume={147},
  number={5},
  pages={4375--4381},
  year={2024},
  doi = {https://doi.org/10.1021/jacs.4c15137},
  publisher={ACS Publications}
}

@article{sanna2024prediction,
  title={{Prediction of ambient pressure conventional superconductivity above \SI{80}{K} in hydride compounds}},
  author={Sanna, Antonio and Cerqueira, Tiago FT and Fang, Yue-Wen and Errea, Ion and Ludwig, Alfred and Marques, Miguel AL},
  journal={npj Comput. Mater.},
  volume={10},
  number={1},
  pages={44},
  year={2024},
  doi={https://doi.org/10.1038/s41524-024-01214-9},
  publisher={Nature Publishing Group UK London},
}

@article{zheng2024prediction,
  title={{Prediction of ambient pressure superconductivity in cubic ternary hydrides with MH6 octahedra}},
  author={Zheng, Feng and Zhang, Zhen and Wu, Zepeng and Wu, Shunqing and Lin, Qiubao and Wang, Renhai and Fang, Yimei and Wang, Cai-Zhuang and Antropov, Vladimir and Sun, Yang and others},
  journal={Mater. Today Phys.},
  volume={42},
  pages={101374},
  year={2024},
  doi={https://doi.org/10.1016/j.mtphys.2024.101374},
  publisher={Elsevier}
}

@article{dolui2024feasible,
  title={{Feasible route to high-temperature ambient-pressure hydride superconductivity}},
  author={Dolui, Kapildeb and Conway, Lewis J and Heil, Christoph and Strobel, Timothy A and Prasankumar, Rohit P and Pickard, Chris J},
  journal={Phys. Rev. Lett.},
  volume={132},
  number={16},
  pages={166001},
  year={2024},
  doi={https://doi.org/10.1103/PhysRevLett.132.166001},
  publisher={APS}
}

@article{semenok2021superconductivity,
  title={{Superconductivity at \SI{253}{K} in lanthanum--yttrium ternary hydrides}},
  author={Semenok, Dmitrii V and Troyan, Ivan A and Ivanova, Anna G and Kvashnin, Alexander G and Kruglov, Ivan A and Hanfland, Michael and Sadakov, Andrey V and Sobolevskiy, Oleg A and Pervakov, Kirill S and Lyubutin, Igor S and others},
  journal={Mater. Today},
  volume={48},
  pages={18--28},
  year={2021},
  doi={https://doi.org/10.1016/j.mattod.2021.03.025},
  publisher={Elsevier}
}

@article{chen2024synthesis,
  title={{Synthesis and superconductivity in yttrium-cerium hydrides at high pressures}},
  author={Chen, Liu-Cheng and Luo, Tao and Cao, Zi-Yu and Dalladay-Simpson, Philip and Huang, Ge and Peng, Di and Zhang, Li-Li and Gorelli, Federico Aiace and Zhong, Guo-Hua and Lin, Hai-Qing and others},
  journal={Nature Commun.},
  volume={15},
  number={1},
  pages={1809},
  year={2024},
  doi={https://doi.org/10.1038/s41467-024-46133-x},
  publisher={Nature Publishing Group UK London}
}

@article{bi2022giant,
  title={{Giant enhancement of superconducting critical temperature in substitutional alloy \ch{(La, Ce)H9}}},
  author={Bi, Jingkai and Nakamoto, Yuki and Zhang, Peiyu and Shimizu, Katsuya and Zou, Bo and Liu, Hanyu and Zhou, Mi and Liu, Guangtao and Wang, Hongbo and Ma, Yanming},
  journal={Nature Commun.},
  volume={13},
  number={1},
  pages={5952},
  year={2022},
  doi={https://doi.org/10.1038/s41467-022-33743-6},
  publisher={Nature Publishing Group UK London}
}

@Inbook{Goldschmidt1967,
author="Goldschmidt, H. J.",
title="Hydrides",
bookTitle="Interstitial Alloys",
year="1967",
publisher="Springer US",
address="Boston, MA",
pages="445--531",
isbn="978-1-4899-5880-8",
doi="10.1007/978-1-4899-5880-8_9",
url="https://doi.org/10.1007/978-1-4899-5880-8_9"
}

@article{meng2019superconductivity,
  title={{Superconductivity of the hydrogen-rich metal hydride \ch{Li5MoH11} under high pressure}},
  author={Meng, Dezhong and Sakata, Masafumi and Shimizu, Katsuya and Iijima, Yuki and Saitoh, Hiroyuki and Sato, Toyoto and Takagi, Shigeyuki and Orimo, Shin-ichi},
  journal={Phys. Rev. B},
  volume={99},
  number={2},
  pages={024508},
  year={2019},
  doi={https://doi.org/10.1103/PhysRevB.99.024508},
  publisher={APS}
}

@article{muramatsu2015metallization,
  title={{Metallization and superconductivity in the hydrogen-rich ionic salt \ch{BaReH9}}},
  author={Muramatsu, Takaki and Wanene, Wilson K and Somayazulu, Maddury and Vinitsky, Eugene and Chandra, Dhanesh and Strobel, Timothy A and Struzhkin, Viktor V and Hemley, Russell J},
  journal={J. Phys. Chem. C},
  volume={119},
  number={32},
  pages={18007--18013},
  year={2015},
  doi={https://doi.org/10.1021/acs.jpcc.5b03709},
  publisher={ACS Publications}
}

@article{song2023stoichiometric,
  title={{Stoichiometric ternary superhydride \ch{LaBeH8} as a new template for high-temperature superconductivity at \SI{110}{K} under \SI{80}{GPa}}},
  author={Song, Yinggang and Bi, Jingkai and Nakamoto, Yuki and Shimizu, Katsuya and Liu, Hanyu and Zou, Bo and Liu, Guangtao and Wang, Hongbo and Ma, Yanming},
  journal={Phys. Rev. Lett.},
  volume={130},
  number={26},
  pages={266001},
  year={2023},
  doi={https://doi.org/10.1103/PhysRevLett.130.266001},
  publisher={APS}
}

@article{song2024superconductivity,
  title={{Superconductivity above 105 K in nonclathrate ternary lanthanum borohydride below megabar pressure}},
  author={Song, Xiaoxu and Hao, Xiaokuan and Wei, Xudong and He, Xin-Ling and Liu, Hanyu and Ma, Liang and Liu, Guangtao and Wang, Hongbo and Niu, Jingyu and Wang, Shaojie and others},
  journal={J. Am. Chem. Soc.},
  volume={146},
  number={20},
  pages={13797--13804},
  year={2024},
  doi={https://doi.org/10.1021/jacs.3c14205},
  publisher={ACS Publications}
}

@article{cerqueira2024searching,
  title={{Searching materials space for hydride superconductors at ambient pressure}},
  author={Cerqueira, Tiago FT and Fang, Yue-Wen and Errea, Ion and Sanna, Antonio and Marques, Miguel AL},
  journal={Adv. Funct. Mater.},
  volume={34},
  number={40},
  pages={2404043},
  year={2024},
  doi={ https://doi.org/10.1002/adfm.202404043},
  publisher={Wiley Online Library}
}

@article{liu2024high,
  title={{High-Throughput Study of Ambient-Pressure High-Temperature Superconductivity in Ductile Few-Hydrogen Metal-Bonded Perovskites}},
  author={Liu, Shi-ming and Shi, Jun-jie and He, Yong and Tian, Chong and Zhu, Yao-hui and Wang, Xinqiang and Zhong, Hong-xia},
  journal={Adv. Funct. Mater.},
  volume={34},
  number={41},
  pages={2315386},
  year={2024},
  doi={https://doi.org/10.1002/adfm.202315386},
  publisher={Wiley Online Library}
}

@article{li2024theoretical,
  title={{Theoretical prediction of high-temperature superconductivity in \ch{SrAuH3} at ambient pressure}},
  author={Li, Bin and Zhu, Cong and Zhai, Junjie and Yin, Chuanhui and Fan, Yuxiang and Cheng, Jie and Liu, Shengli and Shi, Zhixiang},
  journal={Phys. Rev. B},
  volume={110},
  number={21},
  pages={214504},
  year={2024},
  doi={https://doi.org/10.1103/PhysRevB.110.214504},
  publisher={APS}
}

@article{wines2024data,
  title={{Data-driven design of high pressure hydride superconductors using DFT and deep learning}},
  author={Wines, Daniel and Choudhary, Kamal},
  journal={Mater. Futures},
  volume={3},
  number={2},
  pages={025602},
  year={2024},
  doi={https://doi.org/10.1088/2752-5724/ad4a94},
  publisher={IOP Publishing}
}

@article{quan2025prediction,
  title={{Prediction of ambient-pressure superconductivity in the perovskite hydrides \ch{MScH3} (M = K, Rb, Cs)}},
  author={Quan, Hao and Shi, Xian-Biao and Han, Yu-Lin and Zhang, Ping and Wang, Bao-Tian},
  journal={Phys. Rev. B},
  volume={111},
  number={13},
  pages={134509},
  year={2025},
  doi={https://doi.org/10.1103/PhysRevB.111.134509},
  publisher={APS}
}

@article{cerqueira2024sampling,
  title={{Sampling the materials space for conventional superconducting compounds}},
  author={Cerqueira, Tiago FT and Sanna, Antonio and Marques, Miguel AL},
  journal={Adv. Mater.},
  volume={36},
  number={1},
  pages={2307085},
  year={2024},
  doi={https://doi.org/10.1002/adma.202307085},
  publisher={Wiley Online Library}
}

@article{he2023phonon,
  title={{Phonon-mediated superconductivity in the metal-bonded perovskite \ch{Al4H} up to \SI{54}{K} under ambient pressure}},
  author={He, Yong and Lu, Jing and Wang, Xinqiang and Shi, Jun-jie},
  journal={Phys. Rev. B},
  volume={108},
  number={5},
  pages={054515},
  year={2023},
  doi={https://doi.org/10.1103/PhysRevB.108.054515},
  publisher={APS}
}

@article{toby2013gsas,
  title={{GSAS-II: the genesis of a modern open-source all purpose crystallography software package}},
  author={Toby, Brian H and Von Dreele, Robert B},
  journal={J. Appl. Crystallogr.},
  volume={46},
  number={2},
  pages={544--549},
  year={2013},
  doi={https://doi.org/10.1107/S0021889813003531},
  publisher={International Union of Crystallography}
}

@article{stritzker1972superconductivity,
  title={{Superconductivity in the palladium-hydrogen and the palladium-deuterium systems}},
  author={Stritzker, B and Buckel, W},
  journal={Z. Phys. A},
  volume={257},
  number={1},
  pages={1--8},
  year={1972},
  doi={https://doi.org/10.1007/BF01398191},
  publisher={Springer}
}

@article{schirber1974concentration,
  title={{Concentration dependence of the superconducting transition temperature in \ch{PdHx} and \ch{PdDx}}},
  author={Schirber, JE and Northrup Jr, CJM},
  journal={Phys. Rev. B},
  volume={10},
  number={9},
  pages={3818},
  year={1974},
  doi={https://doi.org/10.1103/PhysRevB.10.3818},
  publisher={APS}
}

@article{bronger1994synthese,
  title={{Synthese und Struktur von $\mathrm{SrPdH_{2.7}}$}},
  author={Bronger, W and Ridder, G},
  journal={J. Alloy. Compd.},
  volume={210},
  number={1-2},
  pages={53--55},
  year={1994},
  publisher={Elsevier},
  doi={https://doi.org/10.1016/0925-8388(94)90114-7},
}

@article{olofsson1999novel,
  title={{A novel tetrahedral formally zerovalent-palladium hydrido complex stabilized by divalent alkaline earth counterions}},
  author={Olofsson-M{\aa}rtensson, Malin and Kritikos, Mikael and Nor{\'e}us, Dag},
  journal={J. Am. Chem. Soc.},
  volume={121},
  number={47},
  pages={10908--10912},
  year={1999},
  publisher={ACS Publications},
  doi={https://doi.org/10.1021/ja991047r},
}

@article{stanitski1972ternary,
  title={{Ternary hydrides of calcium and strontium with palladium}},
  author={Stanitski, Conrad and Tanaka, John},
  journal={J. Solid State Chem.},
  volume={4},
  number={3},
  pages={331--339},
  year={1972},
  publisher={Elsevier}
}

@article{yagyu2013synthesis,
  title={{Synthesis of perovskite-type hydrides \ch{APdH3} (A = Sr, Ba) by a new method using \ch{CaH2} as a \ch{H2}-source}},
  author={Yagyu, H and Kato, M and Noji, T and Koike, Y},
  journal={Phys. Procedia},
  volume={45},
  pages={109--112},
  year={2013},
  publisher={Elsevier},
  doi={https://doi.org/10.1016/j.phpro.2013.04.06},
}

@article{errea2013first,
  title={{First-principles theory of anharmonicity and the inverse isotope effect in superconducting palladium-hydride compounds}},
  author={Errea, Ion and Calandra, Matteo and Mauri, Francesco},
  journal={Phys. Rev. Lett.},
  volume={111},
  number={17},
  pages={177002},
  year={2013},
  publisher={APS},
  doi={https://doi.org/10.1103/PhysRevLett.111.177002},
}

@article{kogler2025isome,
  title={{IsoME: Streamlining High-Precision Eliashberg Calculations}},
  author={Kogler, Eva and Spath, Dominik and Lucrezi, Roman and Mori, Hitoshi and Zhu, Zien and Li, Zhenglu and Margine, Elena R and Heil, Christoph},
  journal={Comput. Phys. Commun.},
  pages={109720},
  year={2025},
  publisher={Elsevier},
  doi={https://doi.org/10.1016/j.cpc.2025.109720},
}

@article{lee2023electron,
  title={{Electron--phonon physics from first principles using the EPW code}},
  author={Lee, Hyungjun and Ponc{\'e}, Samuel and Bushick, Kyle and Hajinazar, Samad and Lafuente-Bartolome, Jon and Leveillee, Joshua and Lian, Chao and Lihm, Jae-Mo and Macheda, Francesco and Mori, Hitoshi and others},
  journal={npj Comput. Mater.},
  volume={9},
  number={1},
  pages={156},
  year={2023},
  publisher={Nature Publishing Group UK London},
  doi={https://doi.org/10.1038/s41524-023-01107-3},
}

@article{morel1962a,
  title = {{Calculation of the Superconducting State Parameters with Retarded Electron-Phonon Interaction}},
  author = {Morel, P. and Anderson, P. W.},
  journal = {Phys. Rev.},
  volume = {125},
  issue = {4},
  pages = {1263--1271},
  numpages = {0},
  year = {1962},
  month = {Feb},
  publisher = {American Physical Society},
  doi = {10.1103/PhysRev.125.1263},
  url = {https://link.aps.org/doi/10.1103/PhysRev.125.1263}
}

@article{lucrezi2024full,
  title={{Full-bandwidth anisotropic Migdal-Eliashberg theory and its application to superhydrides}},
  author={Lucrezi, Roman and Ferreira, Pedro P and Hajinazar, Samad and Mori, Hitoshi and Paudyal, Hari and Margine, Elena R and Heil, Christoph},
  journal={Commun. Phys.},
  volume={7},
  number={1},
  pages={33},
  year={2024},
  publisher={Nature Publishing Group UK London},
  doi={https://doi.org/10.1038/s42005-024-01528-6},
}

@article{giannozzi2017advanced,
  title={{Advanced capabilities for materials modelling with QUANTUM ESPRESSO}},
  author={Giannozzi, Paolo and Andreussi, Oliviero and Brumme, Thomas and Bunau, Oana and Nardelli, M Buongiorno and Calandra, Matteo and Car, Roberto and Cavazzoni, Carlo and Ceresoli, Davide and Cococcioni, Matteo and others},
  journal={J. Phys.: Condens. Matter},
  volume={29},
  number={46},
  pages={465901},
  year={2017},
  publisher={IOP Publishing},
  doi={10.1088/1361-648X/aa8f79},
}

@article{migdal1958,
  title={{Interaction between electrons and lattice vibrations in a normal metal}},
  author={Migdal, AB},
  journal={Sov. Phys. JETP},
  volume={7},
  number={6},
  pages={996--1001},
  year={1958},
  doi={http://83.149.229.155/cgi-bin/dn/e_007_06_0996.pdf},
}

@article{eliashberg1960,
  title={{Interactions between electrons and lattice vibrations in a superconductor}},
  author={Eliashberg, GM},
  journal={Sov. Phys. JETP},
  volume={11},
  number={3},
  pages={696--702},
  year={1960},
  doi={}
}

@article{AllenDynes1975,
  title = {{Transition temperature of strong-coupled superconductors reanalyzed}},
  author = {Allen, P. B. and Dynes, R. C.},
  journal = {Phys. Rev. B},
  volume = {12},
  issue = {3},
  pages = {905--922},
  numpages = {0},
  year = {1975},
  month = {Aug},
  publisher = {American Physical Society},
  doi = {10.1103/PhysRevB.12.905},
}

@article{Pellegrini2024,
author = {Pellegrini, Camilla and Sanna, Antonio},
year = {2024},
month = {07},
pages = {},
title = {{Ab initio methods for superconductivity}},
volume = {6},
journal = {Nat. Rev. Phys.},
doi = {10.1038/s42254-024-00738-9}
}

@article{Werthamer_1970,
  title = {{Self-Consistent Phonon Formulation of Anharmonic Lattice Dynamics}},
  author = {Werthamer, N. R.},
  journal = {Phys. Rev. B},
  volume = {1},
  issue = {2},
  pages = {572--581},
  numpages = {0},
  year = {1970},
  month = {Jan},
  publisher = {American Physical Society},
  doi = {10.1103/PhysRevB.1.572},
}

@article{Monacelli_2021,
	doi = {10.1088/1361-648x/ac066b},
	year = 2021,
	month = {jul},
	publisher = {{IOP} Publishing},
	volume = {33},
	number = {36},
	pages = {363001},
	author = {Lorenzo Monacelli and Raffaello Bianco and Marco Cherubini and Matteo Calandra and Ion Errea and Francesco Mauri},
	title = {{The stochastic self-consistent harmonic approximation: calculating vibrational properties of materials with full quantum and anharmonic effects}},
	journal = {J. Phys.: Condens. Matter}
}

@article{baroni2001phonons,
  title={{Phonons and related crystal properties from density-functional perturbation theory}},
  author={Baroni, Stefano and De Gironcoli, Stefano and Dal Corso, Andrea and Giannozzi, Paolo},
  journal={Rev. Mod. Phys.},
  volume={73},
  number={2},
  pages={515},
  year={2001},
  publisher={APS}
}

@article{lucrezi2023quantum,
  title={{Quantum lattice dynamics and their importance in ternary superhydride clathrates}},
  author={Lucrezi, Roman and Kogler, Eva and Di Cataldo, Simone and Aichhorn, Markus and Boeri, Lilia and Heil, Christoph},
  journal={Commun. Phys.},
  volume={6},
  number={1},
  pages={298},
  year={2023},
  publisher={Nature Publishing Group UK London},
  doi={https://doi.org/10.1038/s42005-023-01413-8},
}

@article{lucrezi2024,
  title={{Temperature and quantum anharmonic lattice effects on stability and superconductivity in lutetium trihydride}},
  author={Lucrezi, Roman and Ferreira, Pedro P and Aichhorn, Markus and Heil, Christoph},
  journal={Nature Commun.},
  volume={15},
  number={1},
  pages={441},
  year={2024},
  publisher={Nature Publishing Group UK London},
  doi={10.1038/s41467-023-44326-4}
}

@article{bardeen1957theory,
  title={{Theory of superconductivity}},
  author={Bardeen, John and Cooper, Leon N and Schrieffer, John Robert},
  journal={Phys. Rev.},
  volume={108},
  number={5},
  pages={1175},
  year={1957},
  publisher={APS},
  doi={https://doi.org/10.1103/PhysRev.108.1175},
}

@article{ferreira2024,
title = {{Ab initio modeling of superconducting alloys}},
journal = {Mater. Today Phys.},
volume = {48},
pages = {101547},
year = {2024},
issn = {2542-5293},
doi = {https://doi.org/10.1016/j.mtphys.2024.101547},
author = {P.N. Ferreira and R. Lucrezi and I. Guilhon and M. Marques and L.K. Teles and C. Heil and L.T.F. Eleno},
}

@misc{semenok2025,
      title={{Stability and Superconductivity of Ternary Polyhydrides}}, 
      author={Dmitrii V. Semenok and Di Zhou and Wuhao Chen and Alexander G. Kvashnin and Andrey V. Sadakov and Toni Helm and Pedro N. Ferreira and Christoph Heil and Vladimir M. Pudalov and Ivan A. Troyan and Viktor V. Struzhkin},
      year={2025},
      eprint={2509.22877},
      archivePrefix={arXiv},
      primaryClass={cond-mat.supr-con},
      url={https://arxiv.org/abs/2509.22877}, 
}

@article{QE1,
  title={{QUANTUM ESPRESSO: a modular and open-source software project for quantum simulations of materials}},
  author={Giannozzi, Paolo and Baroni, Stefano and Bonini, Nicola and Calandra, Matteo and Car, Roberto and Cavazzoni, Carlo and Ceresoli, Davide and Chiarotti, Guido L and Cococcioni, Matteo and Dabo, Ismaila and others},
  journal={J. Phys.: Condens. Matter},
  volume={21},
  number={39},
  pages={395502},
  year={2009},
  publisher={IOP Publishing},
  doi={10.1088/0953-8984/21/39/395502}
}

@article{QE3,
    author = {Giannozzi, Paolo and Baseggio, Oscar and Bonfà, Pietro and Brunato, Davide and Car, Roberto and Carnimeo, Ivan and Cavazzoni, Carlo and de Gironcoli, Stefano and Delugas, Pietro and Ferrari Ruffino, Fabrizio and Ferretti, Andrea and Marzari, Nicola and Timrov, Iurii and Urru, Andrea and Baroni, Stefano},
    title = {{QUANTUM ESPRESSO toward the exascale}},
    journal = {J. Chem. Phys.},
    volume = {152},
    number = {15},
    pages = {154105},
    year = {2020},
    month = {04},
    issn = {0021-9606},
    doi = {10.1063/5.0005082},
}

@article{ONCV1,
  title = {{Optimized norm-conserving Vanderbilt pseudopotentials}},
  author = {Hamann, D. R.},
  journal = {Phys. Rev. B},
  volume = {88},
  issue = {8},
  pages = {085117},
  numpages = {10},
  year = {2013},
  month = {Aug},
  publisher = {American Physical Society},
  doi = {10.1103/PhysRevB.88.085117},
}

@article{ONCV2,
title = {{Optimization algorithm for the generation of ONCV pseudopotentials}},
journal = {Comput. Phys. Commun.},
volume = {196},
pages = {36-44},
year = {2015},
issn = {0010-4655},
doi = {https://doi.org/10.1016/j.cpc.2015.05.011},
author = {Martin Schlipf and François Gygi},
}

@article{PBE,
  title = {{Generalized Gradient Approximation Made Simple}},
  author = {Perdew, John P. and Burke, Kieron and Ernzerhof, Matthias},
  journal = {Phys. Rev. Lett.},
  volume = {77},
  issue = {18},
  pages = {3865--3868},
  numpages = {0},
  year = {1996},
  month = {Oct},
  publisher = {American Physical Society},
  doi = {10.1103/PhysRevLett.77.3865},
}

@article{MPsampling,
  title = {{Special points for Brillouin-zone integrations}},
  author = {Monkhorst, Hendrik J. and Pack, James D.},
  journal = {Phys. Rev. B},
  volume = {13},
  issue = {12},
  pages = {5188--5192},
  numpages = {0},
  year = {1976},
  month = {Jun},
  publisher = {American Physical Society},
  doi = {10.1103/PhysRevB.13.5188},
}

@article{MP-smearing,
  title = {{High-precision sampling for Brillouin-zone integration in metals}},
  author = {Methfessel, M. and Paxton, A. T.},
  journal = {Phys. Rev. B},
  volume = {40},
  issue = {6},
  pages = {3616--3621},
  numpages = {0},
  year = {1989},
  month = {Aug},
  publisher = {American Physical Society},
  doi = {10.1103/PhysRevB.40.3616},
}

@article{DFPT,
  title = {{Phonons and related crystal properties from density-functional perturbation theory}},
  author = {Baroni, Stefano and de Gironcoli, Stefano and Dal Corso, Andrea and Giannozzi, Paolo},
  journal = {Rev. Mod. Phys.},
  volume = {73},
  issue = {2},
  pages = {515--562},
  numpages = {0},
  year = {2001},
  month = {Jul},
  publisher = {American Physical Society},
  doi = {https://doi.org/10.1103/RevModPhys.73.515},
}

@article{DFT,
  title = {{Inhomogeneous Electron Gas}},
  author = {Hohenberg, P. and Kohn, W.},
  journal = {Phys. Rev.},
  volume = {136},
  issue = {3B},
  pages = {B864--B871},
  numpages = {0},
  year = {1964},
  month = {Nov},
  publisher = {American Physical Society},
  doi = {10.1103/PhysRev.136.B864},
}

@article{Kohn-Sham,
  title = {{Self-Consistent Equations Including Exchange and Correlation Effects}},
  author = {Kohn, W. and Sham, L. J.},
  journal = {Phys. Rev.},
  volume = {140},
  issue = {4A},
  pages = {A1133--A1138},
  numpages = {0},
  year = {1965},
  month = {Nov},
  publisher = {American Physical Society},
  doi = {10.1103/PhysRev.140.A1133},
}

@Article{Shapeev2015-MTP,
  author = {Shapeev, Alexander V.},
  title = {{Moment Tensor Potentials: A Class of Systematically Improvable Interatomic Potentials}},
  journal = {Multiscale Model. Sim.},
  year = {2016},
  volume = {14},
  pages = {1153-1173},
  number = {3},
  doi = {10.1137/15M1054183},
  cat = {IP}
}

@article{Novikov_2021,
doi = {10.1088/2632-2153/abc9fe},
year = {2020},
month = {dec},
publisher = {IOP Publishing},
volume = {2},
number = {2},
pages = {025002},
author = {Ivan S Novikov and Konstantin Gubaev and Evgeny V Podryabinkin and Alexander V Shapeev},
title = {{The {MLIP} package: moment tensor potentials with {MPI} and active learning}},
journal = {Mach. Learn.: Sci. Technol.}
}

@article{Hansen_2024,
  title = {{Synthesis of \ch{Mg2ItH5}: A potential pathway to high-${T}_{c}$ hydride superconductivity at ambient pressure}},
  author = {Hansen, Mads F. and Conway, Lewis J. and Dolui, Kapildeb and Heil, Christoph and Pickard, Chris J. and Pakhomova, Anna and Mezouar, Mohammed and Kunz, Martin and Prasankumar, Rohit P. and Strobel, Timothy A.},
  journal = {Phys. Rev. B},
  volume = {110},
  issue = {21},
  pages = {214513},
  numpages = {7},
  year = {2024},
  month = {Dec},
  publisher = {American Physical Society},
  doi = {10.1103/PhysRevB.110.214513},
  url = {https://link.aps.org/doi/10.1103/PhysRevB.110.214513}
}

@article{Lu_2025,
  title = {{Prediction and synthesis of \ch{Mg4Pt3H6}: A superconducting complex transition metal hydride stabilized at ambient pressure}},
  author = {Lu, Wencheng and Hutcheon, Michael J. and Hansen, Mads F. and Dolui, Kapildeb and Sinha, Shubham and Sahoo, Mihir R. and Pickard, Chris J. and Heil, Christoph and Pakhomova, Anna and Mezouar, Mohamed and Daisenberger, Dominik and Chariton, Stella and Prakapenka, Vitali and Julian, Matthew N. and Prasankumar, Rohit P. and Strobel, Timothy A.},
  journal = {Phys. Rev. B},
  volume = {112},
  issue = {9},
  pages = {094513},
  numpages = {9},
  year = {2025},
  month = {Sep},
  publisher = {American Physical Society},
  doi = {10.1103/hkx1-lytx},
  url = {https://link.aps.org/doi/10.1103/hkx1-lytx}
}

@article{Garrity2014,
  title={{Pseudopotentials for high-throughput DFT calculations}},
  author={Garrity, Kevin F and Bennett, Joseph W and Rabe, Karin M and Vanderbilt, David},
  journal={Comput. Mater. Sci.},
  volume={81},
  pages={446--452},
  year={2014},
  publisher={Elsevier},
  doi={https://doi.org/10.1016/j.commatsci.2013.08.053},
}

@article{Merchant2023,
  title={Scaling deep learning for materials discovery},
  author={Merchant, Amil and Batzner, Simon and Schoenholz, Samuel S and Aykol, Muratahan and Cheon, Gowoon and Cubuk, Ekin Dogus},
  journal={Nature},
  volume={624},
  number={7990},
  pages={80--85},
  year={2023},
  publisher={Nature Publishing Group UK London},
  doi={https://doi.org/10.1038/s41586-023-06735-9},
}

\end{document}